\documentclass[12pt]{article}
\usepackage{amsmath,amsfonts,amssymb,amsthm}
\usepackage{xcolor}
\usepackage{enumitem}
\usepackage[toc, page]{appendix}
\RequirePackage[colorlinks=true,linkcolor=blue,citecolor=blue,urlcolor=blue,bookmarks,breaklinks=true]{hyperref}

\usepackage{mathtools}

\usepackage{soul,xcolor}
\setul{}{1.2pt}
\usepackage[normalem]{ulem}
\setstcolor{red}

\begin{document}

\title{Transparency versus Anderson localization in one-dimensional
  disordered stealthy hyperuniform layered media}

\renewcommand{\thefootnote}{\fnsymbol{footnote}}
\author{Michael A. Klatt\footnote{German Aerospace Center (DLR), Institute for AI Safety and Security, Wilhelm-Runge-Str. 10, 89081 Ulm, Germany;
German Aerospace Center (DLR), Institute of Frontier Materials on Earth and in Space, Functional, Granular, and Composite Materials, 51170 Cologne, Germany;
Department of Physics, Ludwig-Maximilians-Universität München, Schellingstr. 4, 80799 Munich, Germany;
E-mail: michael.klatt@dlr.de;},
Paul J. Steinhardt\footnote{Department of Physics, Princeton University, Princeton, New Jersey 08544, USA; 
E-mail: steinh@princeton.edu;},
Salvatore Torquato\footnote{Department of Chemistry, Princeton University, Princeton, New Jersey 08544, USA; 
Department of Physics, Princeton University, Princeton, New Jersey 08544, USA; 
Princeton Materials Institute, Princeton University, Princeton, New Jersey 08544, USA; 
Program in Applied and Computational Mathematics, Princeton University, Princeton, New Jersey 08544, USA; 
E-mail: torquato@princeton.edu;}
}

\date{\today}
\maketitle

\begin{abstract}
\noindent
We present numerical simulations of disordered stealthy hyperuniform
layered media ranging up to 10,000 thin slabs of high-dielectric
constant separated by intervals of low dielectric constant  that show
no apparent evidence of Anderson localization of electromagnetic  waves
or deviations from transparency for a continuous band of frequencies
ranging from zero up to some  value $\omega_T$. The results are
consistent with the strong-contrast formula in \cite{kim_effective_2023}
including its tight upper bound on $\omega_T$ and with previous
simulations on much smaller systems. We utilize a transfer matrix method
to compute the Lyaponov exponents, which we show is a more reliable
method for detecting Anderson localization by applying it to a range of
systems with common types of disorder known to exhibit localization,
such as  perturbed periodic lattices. The Lyaponov exponents for these
systems with ordinary disorder show clear evidence of localization, in
contrast to the cases of perfectly periodically spaced slabs and
disordered stealthy hyperuniform layered systems.    As with any
numerical study, one should be cautious about drawing definitive
conclusions.   There remains the challenge of determining whether
one-dimensional disordered stealthy hyperuniform layered media possess a
finite localization length  on some scale much larger than our already
large system size or, alternatively, are exceptions to the standard
Anderson localization theorems.
\end{abstract}


\section*{Introduction}

Anderson localization is a fundamental phenomenon in which
wave transport of all kinds (fermionic, phononic, or electromagnetic)
is halted due to  multiple scattering that
traps waves within a bounded region ~\cite{anderson_absence_1958,
abrahams_scaling_1979}. According to the standard lore, this
localization effect occurs in three-dimensional (3D) systems only if the
degree of disorder is above a critical threshold; however, for
one-dimensional (1D) systems, even an arbitrarily small amount of
disorder leads to
localization for all frequencies~\cite{sheng_introduction_2006}.
Random dimers are a well-known exception. For example, they are
transparent to light propagation for a discrete set of resonant
frequencies due to constructive interference between identical sites
comprising each dimer pair but not for a continuous band of
frequencies~\cite{dunlap_absence_1990}. 

Motivated by recent theoretical studies
\cite{kim_effective_2023,kim_extraordinary_2024}, we present here
comprehensive numerical studies of localization and transparency for the
case of electromagnetic waves propagating in 1D layered systems
consisting of a disordered stealthy hyperuniform  sequence of thin parallel
slabs with a high dielectric constant separated by intervals with low
dielectric constant. We compute the propagation of electromagnetic waves
at normal incidence through up to 10,000
slabs~\cite{yeh_optical_1988,macleod_thin-film_2010} and find no
apparent evidence of Anderson localization  for
a continuous band of frequencies between zero and some critical value
$\omega_T$,  similar to what we obtain for perfectly periodically spaced
slabs over the same frequency range.  In stark contrast, our equivalent
studies of conventionally disordered layered systems that are neither
stealthy nor hyperuniform exhibit clear evidence of localization and
deviation from transparency even when their degree of disorder is much
lower than the disorder in our stealthy hyperuniform examples. Figure~1 is
a schematic representing our transparency versus localization results
for different types of ordered and disordered layered media.

Disordered stealthy hyperuniform many-particle systems are a special
subclass of hyperuniform amorphous states of matter
\cite{torquato_local_2003} with an array of physical properties that
cannot be achieved by nonstealthy hyperuniform or by nonhyperuniform
disordered media \cite{torquato_hyperuniform_2018}. They are defined by
a structure factor $S({\bf k})$ that is exactly zero for a continuous
band of wavenumbers $k \equiv |{\bf k}|$ around the origin ($0 < k < K$)
\cite{batten_classical_2008}, like a crystal, and yet exhibit diffuse
(isotropic) scattering for $k >K$, unlike a crystal
\cite{torquato_hyperuniform_2018}. The fact that no single scattering
can occur from infinite to intermediate wavelengths (${\cal O}(2\pi/K)$)
imposes a continuum of constraints on the disorder, forcing it to be
highly correlated across the sample and to satisfy the bounded-hole
property ~\cite{zhang_can_2017, ghosh_generalized_2018}, which it shares
with periodic structures. It has been suggested
\cite{torquato_hyperuniform_2018} that the  hybrid
crystal-liquid  nature of disordered stealthy hyperuniform materials
endows them with unique and often optimal properties, including
remarkable photonic  \cite{Fl09b,Le16,
Fr17,Zh19,Ta22,klatt_wave_2022,kim_effective_2023,kim_extraordinary_2024,meek_light_2024-1}
and phononic \cite{Gk17,Ro19,Ki20a} characteristics, transport
properties \cite{Zh16b,To18c}  and mechanical properties \cite{To18c}.

While most of the aforementioned studies examined
the unusual properties of 2D and 3D disordered stealthy hyperuniform
materials, recent theoretical work
\cite{kim_effective_2023,kim_extraordinary_2024} has revealed unexpected
localization properties in such 1D media. Specifically, a quantitative
theoretical formula has been derived in Ref.~\cite{kim_effective_2023}
for the effective dynamic dielectric constant based on a nonlocal
strong-contrast formalism developed in
Ref.~\cite{torquato_nonlocal_2021} for general microstructures across
space dimensions and various structural symmetries. 
Because of the fast-convergence properties of strong-contrast
expansions, lower-order truncations yield closed-form approximate
formulas for the effective dielectric constant that apply well beyond
the quasistatic regime, {\it i.e.}, they accurately
estimate the higher-order contributions in the remainder term for a
broad range of incident wavenumbers. In particular, the resulting formula
truncated at third order predicts that 1D disordered stealthy
hyperuniform layered media have perfect transparency (no Anderson
localization) for a range of frequencies
and  specifies a tight upper bound $\omega_c \gtrsim
\omega_T$~\cite{kim_effective_2023, kim_extraordinary_2024}. The
predictions of the strong-contrast formula were also shown to agree with
finite-difference time-domain (FDTD) simulations for 5000
renditions of small systems consisting of up to 50 high-dielectric
slabs~\cite{kim_extraordinary_2024}.

The numerical results presented here based on up to 100 renditions
of samples with up to 10,000 high-dielectric slabs, a greater range of
non-stealthy disordered structures in the comparison group, and a
wider range of analytical methods make for a much stronger case that
the effects of wave propagation in disordered stealthy hyperuniform
systems are qualitatively different from all other known cases of
disorder and are in remarkable accord with  the  tight upper bound
given in~\cite{kim_effective_2023}. We observe an apparent  transparency regime for
disordered hyperuniform systems even for the largest system sizes and
for strong dielectric contrasts. 

In this transparency regime, our
disordered stealthy hyperuniform samples behave very similar to
perfectly periodically spaced slabs for all frequencies less than
$\omega_T$.  On the other hand, unlike the case of periodically spaced
slabs, our disordered stealthy hyperuniform samples  exhibit a sharp
transition to a localized regime for $\omega > \omega_T$, where
$\omega_T$ is
close to but always strictly below  $\omega_c$, the upper bound
prediction based on the strong-contrast
formalism~\cite{kim_effective_2023}. 
Furthermore, the
dependence of $\omega_T$ on the dielectric contrast, volume fraction and degree of
stealthiness ({\it i.e.}, the value of $\chi$ as defined below) is also in
very good agreement with the theoretical predictions for $\omega_c$ in
\cite{kim_effective_2023}.
The rapid
transition from transparency to localization is in stark contrast to the
case of {\it perturbed} (periodic) lattices which exhibit no
transparency regime and a localization length less than the sample
size.

\begin{figure}[p]
\centering
\includegraphics[width=\linewidth]{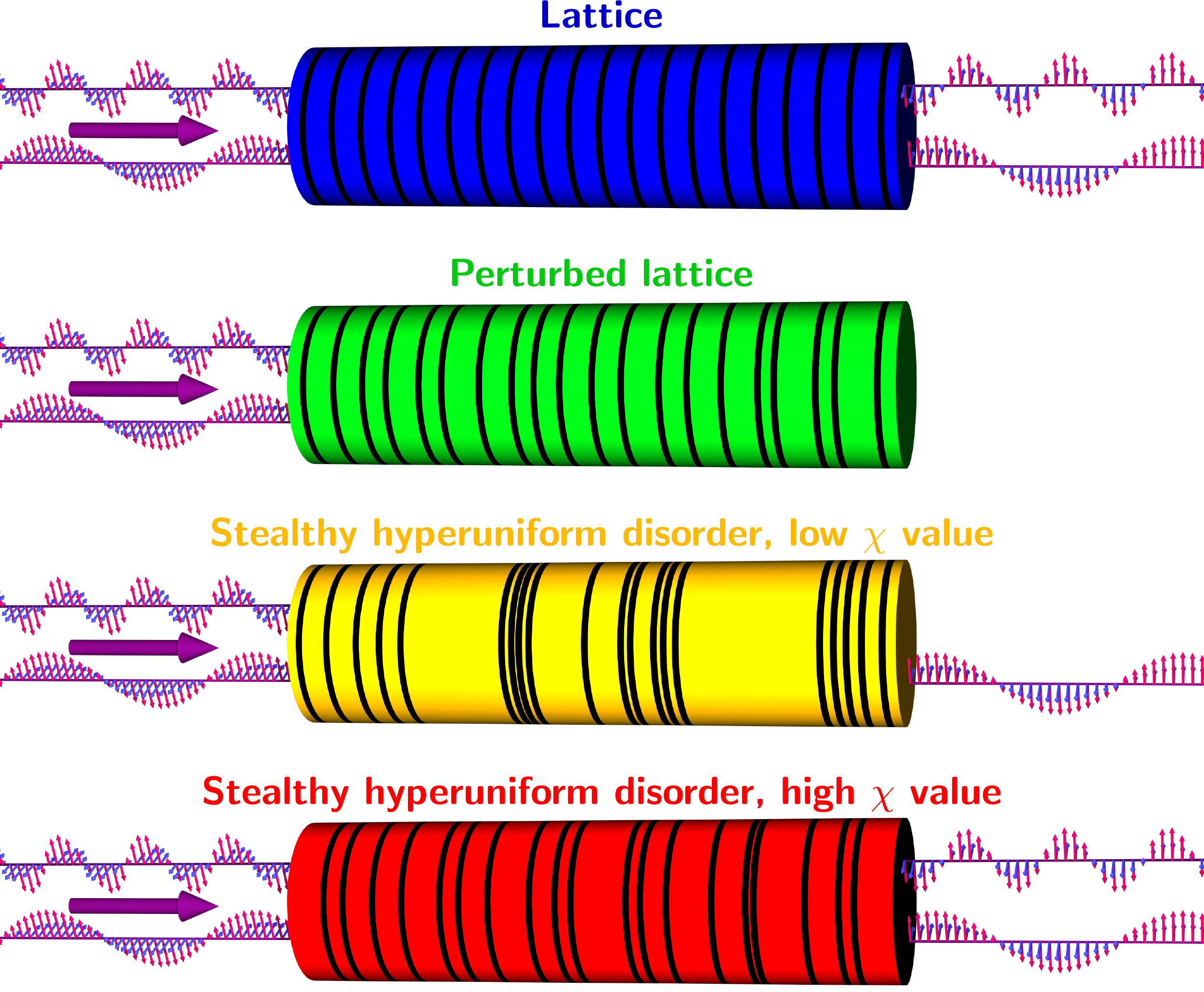}
\caption{Schematic of transparency vs Anderson localization in
layered media that consist of thin high-dielectric slabs (represented by
black layers) and a low-dielectric material in between (represented by
colored layers). Different colors are used for different models: Whereas
a lattice is transparent to waves at most frequencies (top panel), even
a slight perturbation leads to localization and hence no transport
(second panel). A similar schematic for equiluminous or RSA would show
the same outcome. In contrast, disordered stealthy hyperuniform media
have a high degree of local disorder, but they are transparent even for
large system sizes for frequencies $\omega< \omega_T$, where $\omega_T$
is directly proportional to the magnitude of $\chi$, a dimensionless
measure of the range of the wavenumbers in which single scattering in
the two-phase medium is completely forbidden.  Hence, the disordered
stealthy hyperuniform pattern with high $\chi$ (fourth panel) will be
transparent to high  frequencies which are localized for low $\chi$
(third panel). The schematic is consistent with the numerical results
shown in Fig.~2.}
\label{fig:schematic}
\end{figure}

\begin{table*}[b]
\centering
\caption{Standard deviation $\sigma$ of the intervals between high-dielectric slabs
  as a natural metric for local disorder in our models, and 
  the $\tau$-order metric for order across all scales evaluated at the
  maximal system size of $L=10,000$.}
\begin{tabular}{c | c c c c c c c}
  \hline
  \hline\\[-7pt]
  Model                    &
  \multicolumn{3}{c}{Stealthy hyperuniformity}  
                           & 
  Equiluminous             & 
  RSA                      & 
  Perturbed lattice        & 
  Lattice                  \\[2pt]
              &
  $\chi=0.1$  & 
  $\chi=0.2$  & 
  $\chi=0.3$  & 
              & 
              & 
  $a=0.1$     & 
              \\[2pt]
  \hline
  \rule{0pt}{3ex}
  $\sigma$                &
  1.08                    &     
  0.793                   &     
  0.637                   &     
  0.600                   &     
  0.600                   &     
  0.0816\ldots            &     
  0                    \\[2pt]
  $\tau(L)$               &
  3.5                     &     
  3.7                     &     
  4.5                     &     
  0.89                    &     
  1.15                    &     
  23,332                  &     
  $\infty$                    \\[2pt]
  \hline
  \hline
\end{tabular}
\label{tab:order}
\end{table*}

We pay close attention to limitations due to system size, numerical
accuracy and boundary conditions (especially oscillations known as
ripple effects, which play an important role in the analysis, as
described below). We mitigate these effects using various techniques,
{\it e.g.}, via a QR decomposition, averages with high statistics, and
moving windows. We
conclude that the localization length is much larger than the large
sample sizes that we have studied. Indeed, as is dramatically
illustrated in Figs.~\ref{fig:phase-transition}
and~\ref{fig:localization_length_function}, our results for disordered
stealthy hyperuniform  layered media are remarkably similar to the
regime of perfect transparency obtained for layered media with perfectly
periodic spacing which obviously have infinite localization length.
Hence, our findings clearly demonstrate disordered  stealthy
hyperuniform systems are qualitatively different from systems with
ordinary types of disorder. That
said, we must caution that  the strong-contrast formula  is accurate
but approximate and the computational results are limited by numerical
noise.  Hence, one must be cautious in extrapolating our results to
much larger size, dielectric contrast, or volume fraction, as elaborated in
the Discussion section.

We note that there has been related recent work on  localization effects
in disordered stealthy hyperuniform layered media by Park et
al.~\cite{park_deep-subwavelength_2025}, but they did not focus on the
transparency properties. Also, Meek and
Florescu~\cite{meek_light_2024-1}  discussed localization in the context
of photonic band gaps in disordered stealthy hyperuniform layered media
using the inverse participation ratio.

\section*{Models}

Our models are 1D layered media consisting of two-phases:
high-dielectric thin slabs that are embedded within a low-dielectric
material. The models span a broad range of local and global disorder and
order, from ordinary disordered (non-stealthy and non-hyperuniform) to a
perfectly ordered lattice. For each model, we start from a 1D point
process on the line with number density
$\rho$~\cite{chiu_stochastic_2013}. 
We then center at each point high-dielectric slabs with equal thickness
$b$ and fill the spaces between with a low-dielectric material. The
volume fraction occupied by the high-dielectric material is $\phi=\rho
b$. We choose the unit of length such that $\rho =1$, which means that
the thickness $b$ of each slab equals $\phi$. Importantly, we only consider
point processes with a distance $\ge \phi$ between  points to avoid
overlapping slabs. Such overlap would, {\it e.g.}, result in a non-stealthy
 layered medium even if the point process was stealthy
hyperuniform~\cite{torquato_hyperuniform_2018}. The spaces between slabs
may vary depending on the underlying point process. We consider the
following cases:

(1) The \textit{lattice} composed of equally spaced slabs represents perfect
order due to perfect periodicity.

(2) In a \textit{perturbed lattice} with perturbation strength $a$, the
position of each point (and hence high-dielectric slab) is independently
and randomly varied by an amount that is uniformly distributed within
the interval $(-a,a)$. To avoid overlaps, $a<1-\phi$. In our case,
$a=0.1$ and $\phi=1\%$. A perturbed lattice is hyperuniform but not
stealthy~\cite{torquato_local_2003}.

(3) \textit{Random sequential adsorption (RSA)} is non-hyperuniform, and
it guarantees non-overlapping slabs~\cite{torquato_random_2002,
chiu_stochastic_2013, torquato_hyperuniform_2018}.

(4) We create \textit{disordered stealthy hyperuniform} layered media
using the collective-coordinate optimization procedure to obtain
disordered ground states with $\chi <1/3$ from random initial conditions
\cite{batten_classical_2008}.
To avoid overlapping slabs, we modify the standard stealthy potential to
include a soft-core repulsion as in \cite{kim_theoretical_2024,
kim_ultradense_2025}.
The degree of stealthiness for underlying point process is measured by
the $\chi$ value, which is proportional to the ratio of the reciprocal
space volume of wave vectors with constrained values to the total number
of degrees of freedom. In one dimension, $\chi:=K/(2\pi\rho)$, where $K$
is the upper limit on the constrained wavenumbers, {\it i.e.}, $S(k)=0$
for $0<k<K$ \cite{torquato_ensemble_2015}. In practice, due to the
limits of numerical precision, we generate high-quality stealthy samples
with values of $S(k)<10^{-20}$ for $0<k<K$, which corresponds to a
spectral density ${\tilde \chi}_{_V}(k)$ for the resulting stealthy
two-phase medium \cite{torquato_random_2002} that is less than
$10^{-24}$. The spectral density ${\tilde \chi}_{_V}(k)$ for a
distribution of identical nonoverlapping slabs in a matrix is
directly determined by $S(k)$ associated with center points of the
slabs \cite{To16b}; see also the supplementary text for the relevant formulas.

(5) For comparison, we also construct layered media from
\textit{equiluminous} point distributions~\cite{batten_classical_2008}
with a constant positive value of the structure factor $S(k)=10^{-2}$
for $0<k<K$, which corresponds to a spectral density ${\tilde \chi}_{_V}(k)$ of about $10^{-6}$
in the same wavenumber range.

For each model, we simulate samples with 10,000 slabs. For obtaining the
number of samples (mostly 100 per model), see the Methods section.

Table~\ref{tab:order} lists, as a natural metric of local order, the
standard deviations $\sigma$ of the intervals separating the
high-dielectric slabs. Disordered stealthy hyperuniform layers with
$\chi=0.1$ have about the same value of $\sigma$ as the ideal gas. Even
for $\chi=0.3$, we observe the same order of magnitude with
$\sigma=0.637$. For the perturbed lattice, we choose a perturbation
strength of $a=0.1$ such that $\sigma$ is about an order of magnitude
smaller than for  our disordered high-$\chi$ stealthy hyperuniform
samples. That is, our perturbed lattice model has a higher degree of
 order across all length scales as measured by the $\tau$-order
metric~\cite{torquato_ensemble_2015} (see also
Table~\ref{tab:order} and the Supplementary Material) or when compared to
disordered stealthy hyperuniform layered media in the classical sense of
variations in the intervals filled with low-dielectric material.

\section*{Transfer matrix method and Lyaponov exponents}

We compute the wave transport through the dielectric layered media for
all of the models above using the classic transfer matrix method
~\cite{yeh_optical_1988,macleod_thin-film_2010}, which is exact under
standard assumptions such as homogeneous dielectric materials. In our
calculations, we assume loss-free transport, normal incidence, and a
dielectric contrast of $\varepsilon_2/\varepsilon_1$ between the high-
and low-dielectric materials. For the low-dielectric material, we use
$\varepsilon_1 = 1$ as for air or the vacuum.   For the volume fraction
of the high-dielectric slabs, we choose $\phi=1\%$, which allows for
strong local disorder in the stealthy hyperuniform samples.
The unit of time is defined by setting the vacuum speed of light to one.

We begin by evaluating the transmission coefficient $T$; for details on
the setup, see the Methods section. For a sensitive localization
analysis, we then compute the finite approximation of the Lyaponov
exponents for a sample of length $L$~\cite{geist_comparison_1990}:
\begin{align}
\lambda(L) := \left\langle \frac{1}{L} \log \|\Pi(L)\| \right\rangle,
\label{eq:def-lyap-N}
\end{align}
where $\|\Pi(L)\|$ is the spectral norm ({\it i.e.}, largest singular value) of
the product of transfer matrices $\Pi(L)$, and the angular brackets
indicate an ensemble average. To avoid numerical instabilities in the
multiplication of 20,000 matrices, we use a QR-decomposition to compute
the spectral norm~\cite{geist_comparison_1990, scales_lyapunov_1997}.

The Lyaponov exponent is then defined in the limit:
\begin{align}
\lambda := \limsup_{L\to\infty} \lambda(L),
\label{eq:def-lyap-limit}
\end{align}
and the localization length as its inverse, $\xi := 1/\lambda$.  For our
finite system sizes, the limit $\lambda=0$ and hence $\xi=\infty$ cannot
be rigorously evaluated. We can only compare the finite approximations
$\lambda(L)$ and $\xi(L) := 1/\lambda(L)$ to those of the lattice, for
which we know that $\lambda=0$ and hence $\xi=\infty$.

Note that instead of computing the Lyaponov exponents for discrete steps
(namely single layers), we use the length of the sample as a
normalization and can thus compute the exponents for any finite interval
of our samples, {\it i.e.}, we can smoothly vary the system size $L$.

\section*{Results}

\begin{figure}[p]
\centering
\includegraphics[width=\linewidth]{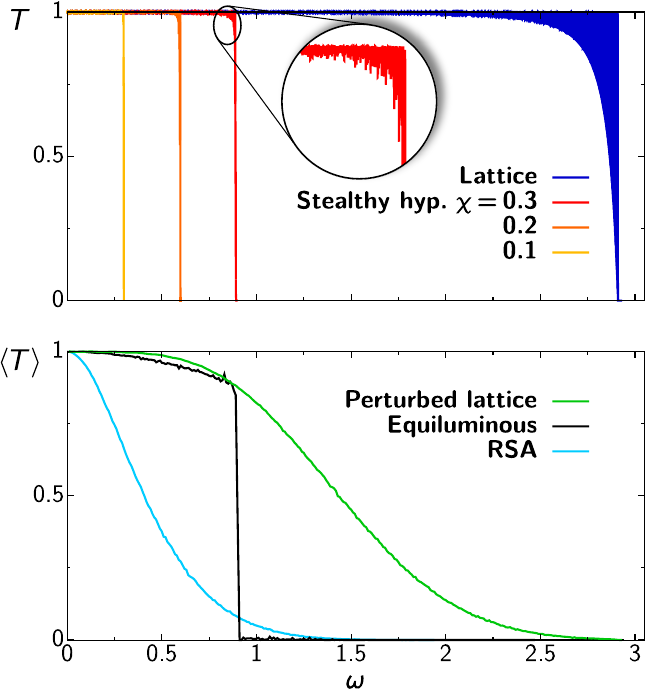}
\caption{Top panel: transmission $T$ through single samples  of a
lattice with periodically spaced slabs (rightmost) compared to
disordered stealthy hyperuniform layered media with different values of
$\chi$  demonstrating that they exhibit  qualitatively similar behavior:
a sharp transition between a transparent and localized regime.
(Here, we only show data up to $\omega_T$ for each case to avoid
confusion with numerical noise in the localized regime at larger
$\omega$.) The transmission for the lattice has very high frequency
oscillations in $T$ (the ripple effect) that appear like a kind of
shading at $\omega \rightarrow \omega_T$ at this resolution;  the inset
shows that the disordered stealthy hyperuniform medium exhibits the same
behavior.  Bottom panel: transmission for layered media with other types
of disorder (perturbed lattice, equiluminous, and RSA) each exhibiting
localization such that $T<1$ for all $\omega > 0$.  The curves are an
average over 100 samples. Each sample in the top and bottom panels
consists of $N=10,000$ slabs with $\varepsilon_2=9$ and $\varepsilon=1$.
We use the same dielectric constants for the incident and substrate medium, respectively.}
\label{fig:transmission}
\end{figure}

We first consider the transmission through a periodic layered medium,
based on the integer lattice; see Fig.~\ref{fig:transmission} (top
panel, blue curve). When numerically resolved, the transmission
coefficient $T$ can be seen to exhibit rapid oscillations caused by the
boundary conditions, an interference phenomenon known as ripple
effect~\cite{macleod_thin-film_2010}. However, $T$ (when numerically
resolved) returns to unity for all frequencies up to a sharp transition
at a critical frequency $\omega_T$, which defines the edge of the first
photonic band gap (PBG)~\cite{joannopoulos_photonic_2008,
dal_negro_waves_2021, yu_engineered_2021}. For our three disordered
stealthy hyperuniform samples with different values of $\chi$, we
observe the same qualitative behavior consistent with the
strong-contrast expansion in \cite{kim_effective_2023}):
apparent transparency up to a
critical frequency $\omega_T$, as shown in the top panel of
Fig.~\ref{fig:transmission}, that includes a ripple effect (seen inset).
Importantly, we observe no dependence  of the transparency regime on the
number of slabs as it ranges from 1000 to 10,000.

This transparency behavior of the perfect lattice and the disordered
stealthy hyperuniform samples in the top panel of
Fig.~\ref{fig:transmission} contrasts sharply with that for layered
media based on disordered but non-stealthy configurations (perturbed
lattice, equiluminous or RSA), as shown in the bottom panel of
Fig.~\ref{fig:transmission}. Comparing a perturbed lattice to RSA, there
is only a quantitative difference in the decay rate (which increases
with system size), but we find the same qualitative behavior, specifically, no sharp transition. For the
equiluminous samples, the behavior is similar to the perturbed lattice
up to $\omega_T$ for the stealthy hyperuniform. 
In no case is there a range of $\omega$ for which  these non-stealthy disordered  media are
transparent.

It is especially notable that one can clearly detect localization for
the perturbed lattice in the bottom panel of Fig.~\ref{fig:transmission}
but not for the disordered stealthy hyperuniform samples shown in the
top panel even though, as mentioned above, the perturbed lattice has
more local order as measured by the $\tau$-order
metric~\cite{torquato_ensemble_2015} or $\sigma$ of the
intervals separating the high-dielectric slabs; see
Table~\ref{tab:order}.

Returning to the top panel of Fig.~\ref{fig:transmission}, the only
obvious difference in the behavior of $T$ for the perfectly periodic
lattice and the disordered stealthy hyperuniform samples is the value of
the threshold $\omega_T(\chi, \varepsilon_2,\varepsilon_1, \phi)$, which
varies with the values of $\chi$, the dielectric contrast and the volume
fraction.
Consistent with  previous theoretical
results~\cite{kim_effective_2023} and small scale simulations from
Ref.~\cite{kim_effective_2023, kim_theoretical_2024}, we find that the
strong-contrast approximate formula provides a reliable upper bound on
the transparency-to-localization threshold $\omega_T$:
\begin{align}
  \omega_T \lesssim \omega_c(\chi; \varepsilon_2,\varepsilon_1, \phi) = \frac{\pi\chi}{\sqrt{\phi(\varepsilon_2-1)+\varepsilon_1}}.
  \label{eq:critical}
\end{align}
This bound is rather tight -- that is, $\omega_T/\omega_c \approx 1$
in all our samples  -- and  becomes sharp in the weak-contrast limit
where $\omega_T/\omega_c \rightarrow 1^-$. (The weak-field behavior is due to the fact that the
nonlocal strong-contrast expansion converges rapidly and is highly
accurate whenever there is coherent wave transport.)

\begin{figure}[p]
  \centering
  \includegraphics[width=\linewidth]{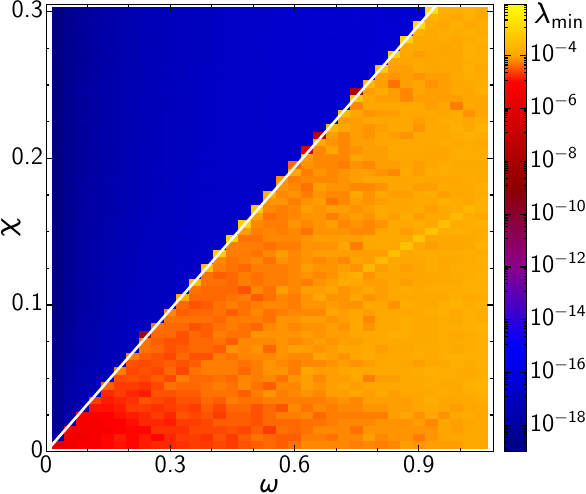}
  \caption{The `phase diagram' of the Lyaponov exponent $\lambda_{\min}=1/\xi$
  for the disordered stealthy hyperuniform samples reveals a sharp 
  transition in agreement with the theoretical prediction in \cite{kim_effective_2023} (solid line).}
  \label{fig:phasediag}
\end{figure}

Since we know that the lattice has a transparency regime with an
infinite localization length and hence a vanishing Lyaponov exponent, we
have to compensate for the ripple-like oscillations in the transmission
plot. For this purpose, we minimize $\lambda(L)$ over a small range of
$L\in[9900,10000]$ that preserves the order of magnitude of the sample.
Since the oscillations may vary with shifts of the samples (due to the
non-periodic setup of the transfer matrix calculations), we use the
median over about 40 shifts per sample for a representative value of
$\lambda_{\min}$, which proves to provide a good approximation for
$\lambda$ for the lattice in the thermodynamic limit; for further
discussions and details on the minimization algorithm, see the Methods
section.

In Figure~\ref{fig:phasediag}, we have systematically studied the
frequency- and $\chi$-dependence of the
transparency-to-localization transition for disordered stealthy
hyperuniform layered media with $\chi<0.3$ to produce a corresponding
`phase diagram' of $\lambda_{\min}(\omega, \chi)$ for
$\varepsilon_2=4$. The figure shows a sharp boundary between a
transparent regime with $\lambda_N<10^{-15}$ (like for a lattice) and a
localized regime with $\lambda_N\approx 10^{-3}$ (like for RSA). The
critical frequency $\omega_c$ as a function of the $\chi$ value is in
good agreement with the strong-contrast formula from \eqref{eq:critical}
(indicated by the solid line).

\begin{figure}[p]
\centering
\includegraphics[width=\linewidth]{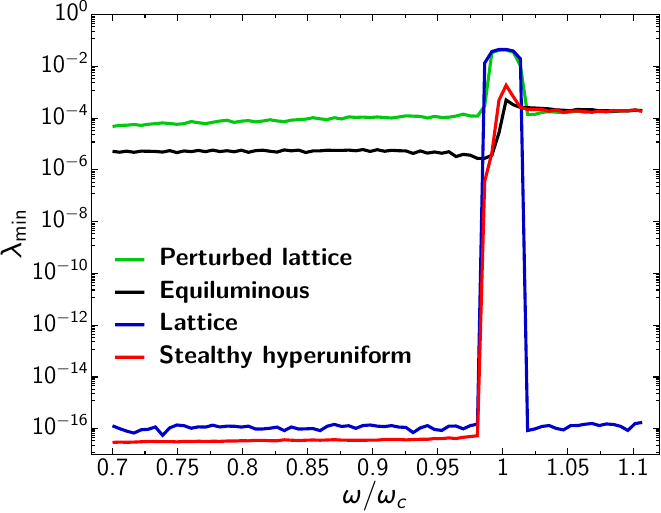}
\caption{Minimal Lyaponov exponent $\lambda_{\min}$ as a function of
  frequency $\omega$ rescaled by $\omega_c$ of the strong-contrast
  formula~\eqref{eq:critical}: as expected, we observe a transparency
  regime for the lattice but localization at all frequencies for the
  perturbed lattice. Surprisingly, the disordered stealthy hyperuniform
  samples exhibit an apparent transparency regime, like for the integer lattice, up
  to $\omega =\omega_T \approx \omega_c$, where we observe a sharp
  transparency-to-localization transition.}
\label{fig:phase-transition}
\end{figure}

\begin{figure*}
\centering
\includegraphics[width=\textwidth]{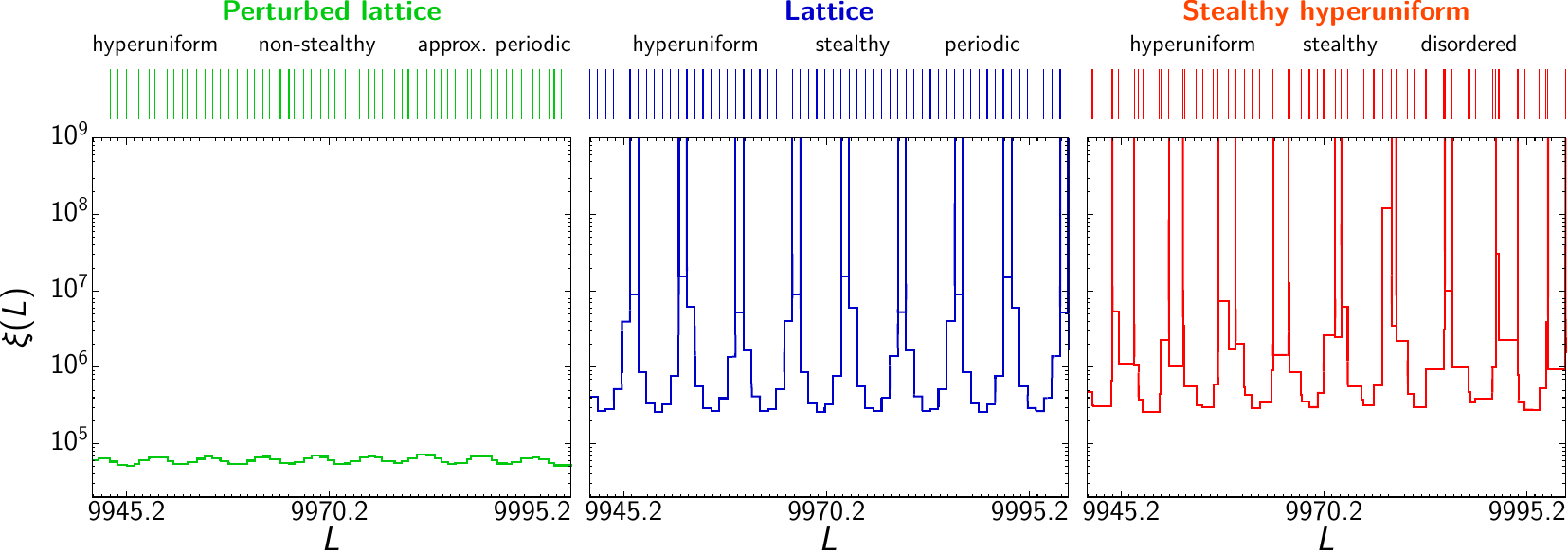}
\caption{Localization length $\xi(L)$ as a function of the system size
  $L$ that varies over about the last 60 slabs for single samples
  using $\varepsilon_2 = 9$, $\phi=1\%$, and
  $\omega=0.5\omega_c$ for $\chi=0.3$;
  corresponding portions of the samples are shown above the plots of
  $\xi(L)$.
  For the perturbed lattice (left), we observe a finite localization
  length for all continuously varied values of $L$. In contrast, for the
  integer lattice (center), the localization length strongly varies due to the
  ripple effect, and within some of the high-dielectric slabs, the
  localization length diverges in sharp peaks. The disordered
  stealthy hyperuniform sample (right) behaves analogously to the lattice. 
  Despite the ripple effect, which causes oscillations, $\xi(L)$ remains larger than
  the system size (with about the same lower values for the lattice and
  the disordered stealthy hyperuniform sample).
  The approximate periodicity in $\xi(L)$ corresponds to $\omega$.
  Recall that $1/\rho$ defines the unit of length.}
\label{fig:localization_length_function}
\end{figure*}

Figure~\ref{fig:phase-transition} provides another way to visualize the
transparency regime for the lattice (with $\chi=1$), where $\lambda_{\min}$ vanishes up
to floating point precision for $\omega < \omega_T \approx 0.98
\omega_c$. 
Remarkably, for the same form of visualization,  the
disordered stealthy hyperuniform samples with $\chi=0.3$ exhibit an apparent transparency
regime that cannot be distinguished numerically from that of the
lattice:  $\lambda_{\min}$ vanishes up to floating point precision.
This stands in stark contradistinction to the standard Anderson
localization lore but in accord with the predictions of the
strong-contrast formula.

As a  comparison, we find  an exponentially larger value of
$\lambda_{\min}$ (or small localization length $\xi=1/\lambda_{\min}$)
at all frequencies for the perturbed lattice  in agreement with the
standard lore of Anderson localization.  The same localization behavior
is found for the equiluminous samples in which $S(k)=10^{-2}$ for
$0<k<K$.
This result for the equiluminous case demonstrates that a modest
suppression of $S(k)$ over the range $0<k<K$ is not sufficient to attain
transparency.

Beyond the transparency regime, i.e., for $\omega > \omega_T$, the 
lattice behaves differently from the stealthy hyperuniform disordered 
medium as is evident in
Fig.~\ref{fig:phase-transition}. The lattice exhibits a PBG
(approximately symmetric around $\omega_c$) followed by a second
transparency regime; for the disordered stealthy hyperuniform samples,
we observe localization for any $\omega > \omega_T$. Put differently, in
contrast to the lattice, the disordered stealthy hyperuniform samples
are fundamentally different  in that they act as a true low-pass
filter across all frequencies.

To understand the transparency regime in more detail, we next consider
$\xi(L)=1/\lambda(L)$ for individual configurations.
Figure~\ref{fig:localization_length_function} shows $\xi(L)$ for about
60 slabs close to $L=10,000$ with $\varepsilon_2 = 9$ and $\phi=1\%$. We
compare the results for a periodic lattice, a perturbed lattice, and a
disordered stealthy hyperuniform layered medium, assuming   the same
input frequency $\omega\approx 0.453$, which corresponds to
$\omega=0.5\omega_c$ for $\chi=0.3$. At the center panel are the results
for the perfect integer lattice, where we know that we are in the
transparency regime.  We observe a complex, approximately periodic
function. Consistent with the ripple effect, the approximate periodicity
of the strong oscillations is set by the frequency $\omega$.
The localization length $\xi(L)$ only changes inside the high-dielectric
phase and remains approximately constant in the low-dielectric phase,
which leads to a `staircase appearance' with rises in the slabs and runs
between.
In certain high-dielectric slabs, $\xi(L)$ diverges in sharp peaks,
consistent with the perfect transparency of the lattice at this
frequency.

On the left hand panel of Fig.~\ref{fig:localization_length_function},
the behavior for the perturbed lattice is qualitatively different. We
find weak oscillations with the same periodicity as for the lattice, but
there are no points of diverging $\xi$ or even finite $\xi \gg L$.
We only find $\xi < L$, i.e,
the localization length is less than the system size for all shifts.
The right  panel shows the results  for the disordered stealthy
hyperuniform sample, where we observe qualitatively the same behavior
as for the lattice. The high-dielectric slabs are strongly disordered
leading to stronger fluctuations in $\xi(L)$ but still with the same
approximate periodicity. For the lattice, we always observe pairs of
sharp peaks in neighboring high-dielectric slabs. For the disordered
stealthy hyperuniform layered media, we also observe pairs of sharp
peaks, but not necessarily in neighboring slabs.

\section*{Discussion}

As noted in the introduction, disordered stealthy hyperuniform layered
media have unusual long-range correlations compared to typical
disordered systems, and, as a consequence, are not subject to Anderson
localization theorems that assume uncorrelated
disorder~\cite{furstenberg_noncommuting_1963, goldsheid_pure_1977}; nor
do they conform to the conditions assumed in theorems that consider
certain kinds of correlated disorder, such as
\cite{aizenman_localization_1993}. Here, we have demonstrated through
highly precise numerical analyses that the disordered stealthy
hyperuniform samples are the only disordered media among those tested
that exhibit an apparent transparency behavior that cannot be
numerically distinguished in Fig.~\ref{fig:phase-transition} from that
of a perfect lattice that is known to be perfectly transparent in the
thermodynamic limit. More precisely, for the accessible large system
size of 10,000 high-dielectric slabs, the disordered stealthy
hyperuniform samples exhibit, within our numerical accuracy, a
continuous transparency regime for $\omega < \omega_T \lesssim \omega_c$
that is in excellent agreement with the strong-contrast formula
in~\cite{kim_effective_2023}. The clear signature of localization
observed for the perturbed lattice and the equiluminous samples in
Fig.~\ref{fig:phase-transition} suggests that relatively small
deviations of any kind from stealthiness destroy perfect transparency.

This observation explains why there are  limitations on how far one can
go in exploring transparency in stealthy hyperuniform systems, whether
crystalline or disordered, since there are inevitable  numerical
inaccuracies.  Increasing the system size, dielectric contrast, or
volume fraction  amplifies the localization effects due to these
inaccuracies.

As a result, the fundamental issue of whether disordered stealthy
hyperuniform media in which the spectral density ${\tilde \chi}_{_V}(k)$
is precisely zero for $0<k<K$ are truly transparent, {\it i.e.}, whether
the localization length is finite or truly infinite, remains open. The
question is equivalent to asking whether the imaginary part of the
remainder term ${\cal R}_4(\omega)$, which embodies all contributions of
fourth- and higher-order terms in the strong-contrast expansion within
the predicted transparency interval
\cite{kim_effective_2023,kim_extraordinary_2024}, is exactly zero or
small but positive in the thermodynamic limit. Resolving the issue is a
theoretical challenge motivated by this study.

At the same time, on a practical level, even if a limiting localization
length is ultimately found, our findings of no apparent evidence for
localization for relatively large system sizes can be used in
applications of photonic and phononic material designs, such as low-pass
filters and thin-films. In any case, the fact that the lattice and
disordered stealthy hyperuniform samples are so similar in the
transmission regime, yet so different beyond, is a tantalizing puzzle to
be explained.

\section*{Materials and Methods}

\paragraph{Samples of layered media}

To compare models with the same number of points per sample (10,000), we
condition our simulation of RSA on the number of disks per sample ({\it i.e.},
we repeat the simulation until the prescribed number of points is
obtained). We simulate the RSA points at a packing fraction of
37\%. Then, we decorate these points with high-dielectric slabs
with a volume fraction of $\phi=1\%$.

We simulate the disordered stealthy hyperuniform samples using the
collective coordinate optimization technique with a soft-core repulsion
as in \cite{kim_theoretical_2024, kim_ultradense_2025}. To preserve a
high-degree of local disorder in our stealthy hyperuniform samples, we
choose a maximum packing fraction of about 1.5\%. We start from
high-temperature initial conditions, more specifically, the ideal gas in
the canonical ensemble. For the optimization of the collective
coordinates, we employ a limited-memory Broyden–Fletcher–
Goldfarb–Shanno (L-BFGS) optimization algorithm. To obtain high-quality
1D stealthy hyperuniform packings in a reasonable simulation time, we
employ a three-step procedure. First, we simulate high-quality samples
without soft-core repulsion with an energy tolerance of $10^{-17}$;
then, we repeat the simulation with soft-core repulsion with a radius of
$0.015$ and an energy tolerance of $10^{-8}$; finally, we simulate
high-quality samples without soft-core repulsion and an energy tolerance
of again $10^{-17}$. Although the last step is without soft-core
repulsion, the remaining small collective displacements still preserve a
volume fraction with $\phi=1\%$ and do not create slab overlaps. For the
equiluminous samples, we used an energy tolerance of $10^{-12}$ for the
steps with hard-core repulsion.

All of our samples are simulated with periodic boundary conditions.
For each model, we simulate at least 100 samples, except for the
lattice where a single sample is representative. For the average
transmission coefficients in Fig.~\ref{fig:transmission} (bottom panel), we
use 10,000 samples per model. For the phase diagram in
Fig.~\ref{fig:phasediag}, we simulate 30 samples for each $\chi$ value
(for each of the 60 different $\chi$ values).

\paragraph{Transmission coefficients}
We use the transfer matrix methods as in \cite{macleod_thin-film_2010}.
For potential pitfalls, in a transmission coefficient analysis of
localization with large system sizes, see Fig.~S1 in the supplementary text.

\paragraph{Localization analysis}
For a numerically stable Lyaponov analysis of localization, we use a
QR-decomposition of the product of transfer matrices as in
\cite{geist_comparison_1990} or later in \cite{scales_lyapunov_1997}
with the help of the Householder
transformation~\cite{eckmann_ruelle_1985}. Based on the lattice
case where we know $T=1$ (or, equivalently, the  localization length
$\xi = 1/\lambda \rightarrow \infty$ as $L\rightarrow \infty$) and what
we find numerically for the lattice for finite $L$,  we infer that
$1/\lambda_{\rm min}$ accurately gauges the localization length when $L$
is large but finite.  We then use  $1/\lambda_{\rm min}$ to gauge the
localization length for the disordered stealthy hyperuniform layered
media, which have such similar behavior to the lattice for $\omega<
\omega_c$, as illustrated in Figures~5 and~6.

Due to the ripple effect that occurs in finite systems, the maximal
numerical value of  $\xi =1/\lambda_{\min}$ is best determined by
varying the range of $L$ over a narrow range near its maximum value,
which in this case is $L=10,000$.  In almost all cases, we have fixed
this range to be  $[9900, 10000]$, except for the case  $\omega<0.1$,
where the long wavelengths in the oscillations require extending the
range to $[9500,10000]$.

We minimize the Lyaponov exponent using a two-step procedure. First, we
identify suitable candidates for sharp peaks (by sufficiently strong
local minima in $\lambda(L)$); then, we iterate a golden-search
minimization routine for each of these candidates (but stop the
iteration if we find a functional value smaller than $10^{-15}$). Note
that just taking the smallest value of a preliminary search leads to
systematic errors because of the strong oscillations in $\lambda(L)$
which may not always correspond to a sharp peak.

Note that the sharp peaks require a fine step size in $L$ (for which we
used 0.0001, where the unit of length is defined via $\rho = 1$).
The oscillations in $\lambda(L)$ vary with shifts of the samples,
which are subject to periodic boundary conditions. The shifts correspond to
independent representations that result in 
different oscillations in $\lambda(L)$ because the product of
transfer matrices does not use periodic boundary conditions. Therefore,
we average over  about 20 shifts for Fig.~\ref{fig:phasediag}. The
shifts should run over multiples of pairs of high-dielectric slabs for
the disordered models. A sufficient number of shifts is required for
good statistics (to not distort the Lyaponov exponent by a single
outlier).
Finally, we compute the average of $\lambda_{min}$ over different
samples.

We then use the median for each single sample to represent the dominant
behavior of $\lambda_{\min}$ since the distribution of the numerically
obtained values of $\lambda_{\min}$ is bimodal even for the perfect
(periodic) lattice. For example,  the median for the lattice
accurately captures the perfect transmission.
More precisely, the values of the Lyaponov exponent are either
$O(10^{-15})$ ({\it i.e.}, zero within floating point precision) or
$O(10^{-6})$ (depending on the system size and physical parameters). The
median represents the dominant behavior and is less sensitive than the
mean value to the precise protocol of our choice of shifts.

\nocite{atkinson, primes, klatt_cloaking_2020, last_lectures_2017}

\subsection*{Acknowledgements}

This research was sponsored by the U.S. Army Research Office and was
accomplished under Cooperative Agreement No. W911NF-22-2-0103.
M.A.K.~acknowledges funding and support by the Initiative and Networking
Fund of the Helmholtz Association through the Project ``DataMat,'' as
well as by the DFG through SPP 2265 under Grant Nos. KL 3391/2-2, WI
5527/1-1, and LO 418/25-1. The simulations presented in this article
were substantially performed on computational resources managed and
supported by the Princeton Institute for Computational Science and
Engineering (PICSciE). The authors also gratefully acknowledge the
scientific support and HPC resources provided by the German Aerospace
Center (DLR). The HPC system CARO is partially funded by ``Ministry of
Science and Culture of Lower Saxony'' and ``Federal Ministry for
Economic Affairs and Climate Action.''

\end{document}